**Universal Scaling Law to Predict the Efficiency of Magnetic Nanoparticles as MRI T2-Contrast Agents**

By
*Quoc L. Vuong, Jean-François Berret, Jérôme Fresnais, Yves Gossuin\* and Olivier Sandre\**

Dr. Q. L. Vuong, Author-One,
Université de Mons, Biological Physics Department, 20 Place du Parc, 7000 Mons, Belgium
Dr. J.-F. Berret, Author-Two,
Université Denis Diderot Paris-VII, CNRS UMR7057, Matière et Systèmes Complexes
10 rue Alice Domon et Léonie Duquet, 75013 Paris, France
Dr. J. Fresnais, Author-Three,
UPMC Univ Paris 06, CNRS UMR7195, Physicochimie, Colloïdes et Sciences Analytiques
4 place Jussieu, 75005 Paris, France
[*]      Dr. Y. Gossuin, Corresponding-Author,
Université de Mons, Biological Physics Department, 20 Place du Parc, 7000 Mons, Belgium
E-mail: yves.gossuin@umons.ac.be
[*]      Dr. O. Sandre, Corresponding-Author,
Université de Bordeaux, CNRS UMR5629, Laboratoire de Chimie des Polymères Organiques
ENSCBP, 16 Avenue Pey Berland, 33607 Pessac, France
E-mail: olivier.sandre@ipb.fr

**Abstract:** Magnetic particles are very efficient Magnetic Resonance Imaging (MRI) contrast agents. In the recent years, chemists have unleashed their imagination to design multi-functional nanoprobes for biomedical applications including MRI contrast enhancement. This study is focused on the direct relationship between the size and magnetization of the particles and their nuclear magnetic resonance relaxation properties, which condition their efficiency. Experimental relaxation results with maghemite particles exhibiting a wide range of sizes and magnetizations are compared to previously published data and to well-established relaxation theories with a good agreement. This allows deriving the experimental master curve of the transverse relaxivity *versus* particle size and to predict the MRI contrast efficiency of any type of magnetic nanoparticles. This prediction only requires the knowledge of the size of the particles impermeable to water protons and the saturation magnetization of the corresponding volume. To predict the $T_2$ relaxation efficiency of magnetic single crystals, the crystal size and magnetization – obtained through a single Langevin fit of a magnetization curve – is the only information needed. For contrast agents made of several magnetic cores assembled into various geometries (dilute fractal aggregates, dense spherical clusters, core-shell micelles, hollow vesicles…), one needs to know a third parameter, namely the intra-aggregate volume fraction occupied by the magnetic materials relatively to the whole (hydrodynamic) sphere. Finally a calculation of the maximum achievable relaxation effect – and the size needed to reach this maximum – is performed for different cases: maghemite single crystals and dense clusters, core-shell particles (oxide layer around a metallic core) and zinc-manganese ferrite crystals.

## 1. Introduction

Magnetic Resonance Imaging contrast agents (MRI CAs) allow a high sensitivity for the early detection of different pathologies and the tracking of magnetically tagged cells *in vivo* through molecular and cellular imaging.[1] The efficiency for MRI CAs consists in lowering the longitudinal ($T_1$) or transverse ($T_2$) relaxation times of the nuclear spins of water protons in tissues at the lowest CA concentration, expressed in equivalent mM of magnetic ions. This acceleration of proton magnetization relaxations near a magnetic particle is usually ascribed to fluctuations of magnetic dipolar interactions between the nuclear and the electronic spins. Therefore the "relaxivity" defined as the slope of the relaxation rate in s$^{-1}$ (either $T_1$ or $T_2$) *versus* the equivalent ion concentration in mM is a direct measurement of this efficiency. In the various magnetic nanoparticles or magnetic hybrids proposed so far, authors have interpreted their relaxivity results compared to literature and specifications of commercial products by referring to different notions such as the nature and the size of the Ultra-small Superparamagnetic Iron Oxide (USPIO) grains, the clustering effect (several USPIOs per contrast agent), the influences of a non magnetic shell, of a hydrophobic membrane, of water permeability… In this work, we propose a general treatment of new results and of literature data with a unified method using only two (for individual USPIOs) or three (for clusters or hybrids) parameters: diameter and magnetization of the whole particle, and volume fraction of magnetic materials inside.

## 2. Magnetic nanoparticles and clusters of various sizes and geometries

### 2.1. Samples prepared for this study

Magnetic particles and clusters were prepared from USPIO nanoparticles made of maghemite ($\gamma$–$Fe_2O_3$) synthesized in water according to Massart's procedure (see Experimental).[2] These nanoparticles were readily dispersed in water and formed a true "ionic ferrofluid". The iron oxide surface bears positive charges due to adsorption of protons in acidic media, in that case a dilute $HNO_3$ solution at pH between 1.2 and 1.7. Such ionic ferrofluids remain in a monophasic state under the application of a magnetic field of arbitrary value. On the microscopic scale, those crystals exhibit a Log-Normal distribution of diameters with parameters $d_0$ =7 nm and $\sigma$ = 0.38, as measured by vibrating sample magnetometry (VSM).[3] Maghemite nanoparticles with such a broad dispersity were treated with a size-sorting procedure based on fractionated phase-separation.[4] After the increase of ionic strength to induce demixtion and magnetic sedimentation on a strong ferrite magnet, a concentrate could be separated from the supernatant in few minutes. By repeating the phase-separation protocol on both the concentrate and the supernatant, we obtained four new fractions at second level of refined distribution of sizes, and so on after a third and fourth level. To take into account the residual polydispersity, we estimated a weight-averaged diameter characteristic of each sample by calculating the 4$^{th}$ and 3$^{rd}$ order moments of the Log-normal distributions deduced by VSM: $d_w = <d^4>/<d^3> = d_0 \exp(3.5\sigma^2)$. This choice of the characteristic size enabled comparing samples of varying size distributions (even for a same median size $d_0$). It also fairly compares to the mean diameter observed on the electron microscopy pictures and to twice the gyration radius that can be obtained by a common experiment (Guinier's plot) from several scattering techniques (light, X-rays or neutrons). After the size-sorting process, the fractions of interest were coated either by coordination bonding of surface iron ions with tri-sodium citrate ($Na_3Cit$) ligand[5] or by electrostatic complexation with sodium polyacrylate ($PAA_{2k}$ or $PAA_{5k}$) using a precipitation-redispersion protocol.[6] Superparamagnetic Iron

Oxide (SPIO) particles consisting in spherical clusters of a limited number of individual crystals were also prepared by a coacervation protocol with oppositely charged hydrophilic copolymers controlled by salinity.[7] Experimental details for the size sorting and coating procedures of the USPIOs and their controlled clustering into SPIO coacervates are described in section 4.

## 2.2. Samples from literature

Among the huge literature published in the recent years on novel particles as potential MRI contrast agents, we focused our attention on articles reporting precisely the physical (structural, geometrical, magnetic…) properties of the systems. Thus we gathered a broad range of data for individually dispersed maghemite nanoparticles with only a thin permeable coating,[8–10] maghemite core-silica shell particles,[11,12] maghemite or magnetite nanoparticles clustered by hydrophilic[13–15] or amphiphilic[16–18] polymers, encapsulated in the aqueous lumen of liposomes[19] or embedded in the membrane of lipid[20] or polymer[21] vesicles. Other magnetic materials such as pristine iron–iron/manganese-ferrite core-shell nanoparticles[22] and manganese-ferrite nanoparticles clusters[23] were also used to complete our relaxation analysis.

Except for Ultra Ultra-small Superparamagnetic Iron Oxide (UUSPIO) of diameter 4 nm only which can be used as positive contrast agents with $T_1$-weighted sequences,[24] magnetic particles are only effective as $T_2$ negative contrast agents since at the magnetic fields of most clinical and research MR imagers (1.5 T or above), their longitudinal relaxation rate $R_1$ falls down due to the absence of the so-called "secular term" in the theoretical expression of $R_1$.[25] Since $R_2/R_1$ becomes very high, any hyper-signal is lost because of spin-spin relaxation in the transverse plane where the detection antennas are sensitive. This was checked for our samples – by recording the evolution of $R_1$ on a large range of magnetic fields – confirming the small effect of the particles on water longitudinal relaxation (**Figure S5** in Supporting Information). On the opposite, water transverse relaxation rate $R_2$ was strongly enhanced by the presence of the samples with no significant variation *versus* magnetic field values between 0.47 and 1.41 T (**Figure S6**).

## 2.3. Classical relaxation models revisited

For magnetic particles of diameter $d$ and saturation magnetization $M_v$ (corresponding to the total magnetic moment divided by the particle volume and expressed in the SI unit, A·m⁻¹), the value of $R_2$ is given either by the motional averaging regime (MAR) – also called outer sphere theory (as opposite to inner sphere) – or by the static dephasing regime (SDR). But to compare the relaxation data with these models, one needs first to properly define the structure and geometry of the $T_2$ contrast agent, as depicted by the different cases on **Scheme 1**. We recall here only the final equations of MAR and SDR models, but readers can refer to recent reviews to learn about the quantum mechanics treatment of magnetic dipolar interactions between nuclear and electronic spins which are founding them.[25]

### 2.3.1. *Particles in the motional average regime (MAR):* $\Delta\omega\tau_D < 1$

In this case, the protons of freely diffusing water molecules surrounding the particle explore all the possible values of magnetic dipolar field created by the electronic magnetic moment and the transverse relaxation rate at high field is given by:

$$R_2 = \frac{1}{T_2} = \frac{16}{45} f \, \tau_D \left(\Delta\omega\right)^2 \tag{1}$$

where $f$ is the volume fraction occupied by the particles in the suspension, $\Delta\omega = \gamma\mu_0 M_v/3$ is the angular frequency shift experienced by a proton at the equator of the particle, $\gamma = 2.67513\times10^8$ rad·s⁻¹·T⁻¹ the gyromagnetic factor of the proton, $\mu_0 = 4\pi10^{-7}$ T·m·A⁻¹ the magnetic permeability of vacuum and $\tau_D = d^2/4D$ the translational diffusion time of the protons in the magnetic field inhomogeneities created by the particles ($D$ being the water translational diffusion constant and $d$ the particle diameter).[25] To obtain the transverse relaxivity $r_2$ – defined as the relaxation rate $R_2$ normalized by the equivalent iron concentration [Fe] in mM – the volume fraction $f$ should be expressed in terms of equivalent iron concentration. This concept of relaxivity is widely used in the literature to normalize the relaxation rates and compare the efficiencies of different contrast agents. The iron concentration is indeed more direct to measure (by UV-Vis absorption, atomic emission or inductively coupled plasma mass spectroscopy) than the volume fraction occupied by the particles in water. However, normalization by the volume fraction is more rigorous from a theoretical point of view as seen on **Equation (1)**. The numerical factor between $f$ and [Fe] depends on the type of magnetic material, through its molar volume $v_{mat}$ given by the ratio of the molar mass divided by the number of magnetic ions in the formula and by the mass density, both expressed in SI units to obtain the iron concentration in mol·m$^{-3}$ equivalent to mmol·L$^{-1}$.

For maghemite ($M_{\gamma\text{-Fe2O3}}$=0.1597 kg·mol$^{-1}$ and $\rho_{\gamma\text{-Fe2O3}}$=5100 kg·m$^{-3}$), this writes:

$$f/[\text{Fe}] = v_{mat} = \frac{M_{\gamma\text{-Fe}_2\text{O}_3}}{2\rho_{\gamma\text{-Fe}_2\text{O}_3}} = 1.57 \times 10^{-5}\,\text{m}^3 \cdot \text{mol}^{-1} \tag{2a}$$

and in the general case of an iron spinel MO·Fe2O3:

$$f/\{[\text{Fe}]+[\text{M}]\} = v_{mat} = \frac{M_{\text{MO·Fe}_2\text{O}_3}}{3\rho_{\text{MO·Fe}_2\text{O}_3}} \approx 1.5 \times 10^{-5}\,\text{m}^3 \cdot \text{mol}^{-1} \tag{2b}$$

The nature of the divalent cation in the iron spinel structure ($Fe^{2+}$, $Co^{2+}$, $Ni^{2+}$, $Mn^{2+}$, $Zn^{2+}$, $\cdots$) varies this result only by ±5%. Then equation (1) writes:

$$r_2 = \frac{R_2}{[\text{Fe}]} = \frac{4\gamma^2 \mu_0^2 v_{mat} M_v^2 d^2}{405 D} \tag{3}$$

which is valid only if the Redfield condition is fulfilled:

$$\Delta\omega\tau_D < 1 \tag{4}$$

This is the case for small single nanoparticles of pure magnetic materials or with a thin fully hydrated shell,[8-10,13] and for hybrid entities with an overall magnetization ($M_v$) remaining small compared to the specific magnetization of the inorganic part ($m_S$), for example dilute micelles[17] or vesicles[19,20] containing few iron oxide nanoparticles, or magnetic cores either single[11] or clustered[12] wrapped by a rather thick silica coating. In the latter case (porous SiO$_2$), the permeability of the non magnetic mantle to water molecules can lead to an additional fast mode, but its contribution to the overall relaxation process remains small as long as the water protons diffusing in the "outer shell" represent a larger volume fraction than the internal protons linked to the porous network. For magnetic particles respecting **Equation (4)**, the transverse relaxivity can be divided by $M_v^2$ to point out the dependence on diameter. Replacing the different constants by their numerical values and using $D = 3 \times 10^{-9}$ m$^2$·s$^{-1}$ as the water diffusion constant at 37 °C leads to:

$$\frac{r_2}{M_v^2} = a_{theo}\, d^2 = 5.9 \times 10^{-12}\, d^2 \tag{5}$$

To test this relationship, the ratio $r_2/M_v^2$ was calculated for 9 sizes of single maghemite USPIOs. The values of saturation magnetization $M_v$ (A·m$^{-1}$) of all these samples were obtained from the fits of the magnetometry curves with a precise knowledge of the solid volume fraction $f$ in the suspension from an independent titration of iron (also necessary to deduce the relaxivities $r_2$ in s$^{-1}$·mM$^{-1}$). Data from literature corresponding to various types of ferrite nanoparticles and clusters in the MAR are also presented on Figure 1. For relaxivities measured at a temperature different from 37 °C, a correction factor corresponding to the tabulated variation of viscosity was applied, reflecting the change of diffusion constant of water molecules.

The values of diameter of the individually dispersed USPIO nanoparticles plotted on the curve are the weight average diameter ($d_w$) obtained by magnetometry (14 samples) or by TEM (4 samples) or from the simulated curve fitting a $T_1$ NMRD profile (3 samples). Concerning the clusters, micelles, vesicles and single magnetic cores surrounded by an impermeable non magnetic coating (*e.g.* silica or polymer), the plotted diameter was either measured by electron microscopy ($d_{TEM}$) or by dynamic light scattering ($d_H$) depending on availability in the corresponding references.

For clusters and hybrids, we introduce the intra-aggregate volume fraction of magnetic materials $\phi_{intra}$ to derive a corrected relaxivity $r_2' = r_2 \times \phi_{intra}$. This normalization for SPIO clusters enables to properly compare their relaxation data $r_2 \times \phi_{intra}/M_v^2$ as if they were filling the same volume fraction of suspension as single USPIO nanoparticles at 1 mM iron concentration. Even though the relaxation efficiency is usually expressed in terms of relaxivity per equivalent mM of iron atoms, the normalization by $M_v^2$ is theoretically justified for the same total volume fraction of particles $f$ in the suspension, including both parts (iron oxide and impermeable coating). For example with clusters containing $\phi_{intra}$ = 10% in volume of magnetic nanoparticles, the volume fraction $f$ of clusters in a 1 mM [Fe] suspension will be 10 times larger than for a 1 mM suspension of the dispersed USPIOs. The measured relaxivities should thus be divided by ten to compare the relaxation efficiencies at the same volume fraction of magnetic particles.

**Figure 1** shows $r_2 \times \phi_{intra}/M_v^2$ versus $d$ for samples following the MAR model (see **Table 1**), from this study and from literature. This figure presents the normalized relaxivity, obtained as described above, while **Figure S7** (supporting information) directly presents the relaxivity in order to ease the comparison with measured values. Both figures show that samples respecting Equation (4) are indeed quantitatively following a quadratic dependence on diameter over almost two decades (4–300 nm):

$$\frac{r_2 \times \phi_{intra}}{M_v^2} = a_{exp} \, d^2 = 11.6 \times 10^{-12} \, d^2 \tag{6}$$

where $a_{exp}$ was obtained by a one-parameter quadratic fit of correlation factor $R$=0.93. The agreement between $a_{exp}$ and $a_{theo}$ values is fairly good, since the size distribution of the nanoparticles – although contained in the weight ($d_w$) or intensity ($d_H$) averaged diameter – is expected to influence relaxation and thus the actual value of prefactor $a$.[26] Moreover, the diffusion coefficient of bulk water was used to calculate prefactor $a_{theo}$ while water diffusion might be hampered in the vicinity of the particles, as shown in the case of especially hydrophilic polymer shells.[27] By using the scaling law expressed in **Equation (6)**, it is now possible to predict the transverse relaxativity at magnetic fields above 1 T and at 37 °C of any sample of particles or clusters inducing relaxation in the MAR whose diameter ($d_w$, $d_{TEM}$ or $d_H$), magnetic content $\phi_{intra}$ and saturation magnetization $M_v$ are known, prior to any NMR or MRI measurement. **Figure 1** also presents the data for two samples whose $\Delta\omega\tau_D$ values are slightly larger than 1 and thus appear below the master curve of MAR, as expected.[10,22] Among these cases, iron-iron oxide core-shell particles[22] locate slightly below the corresponding particles in the MAR with a similar size $d \approx 15$ nm (see **Table 1**). This is logical since for this sample $\Delta\omega\tau_D$ = 1.5, a value corresponding to the transition between MAR and SDR relaxation regimes. Even though $r_2/M_v^2$ is smaller than predicted, the experimental value $r_2$=324 s$^{-1}$ mM$^{-1}$ is twice as large as for a pure iron oxide USPIO of same outer diameter. However, the comparison is somehow complicated by the different stoichiometry and density of both compounds (pristine iron in the core and iron oxide in the shell), implying different volume fractions $f$ for iron oxide and core-shell particles with identical iron concentration: respectively $f=1.57\times10^{-5}$ for iron oxide and $f=10^{-5}$ for Fe@Fe$_2$O$_3$ at 1 mM equivalent [Fe], so that the increase of specific magnetization ($M_v=6.6\times10^5$ A·m$^{-1}$ or 115 emu·g$^{-1}$) is somehow counterbalanced by the decrease of molar volume $v_{mat}$.

### 2.3.2. _Particles out of the motional averaging regime:_ $\Delta\omega\tau_D > 1$

For particles with size $d$ and magnetization $M_v$ such that $\Delta\omega\tau_D > 1$, the Static Dephasing Regime (SDR) – implying that water protons explore only a small space compared to the hydrodynamic volume also called "outer shell" around the particle – should be used instead of MAR, yielding a relaxation rate of the form:

$$R_2^* = \frac{1}{T_2^*} = \frac{2\pi}{3\sqrt{3}}\, f\, \Delta\omega \approx R_2 \tag{7}$$

With the same transformation as above, one obtains:

$$r_2^* = \frac{R_2^*}{[Fe]} = \frac{2\pi\,\gamma\mu_0\, v_{mat}\, M_v}{9\sqrt{3}} \approx r_2 \tag{8}$$

The SDR model does not take into account the effect of the refocusing pulses used in all $T_2$ measurement sequences. Therefore **Equation (7)** and **Equation (8)** are only exact for $R_2^*$ and $r_2^*$ respectively, which determine the signal contrast in MR imaging conditions without spin echoes (_e.g._ with gradient echoes sequences). Nevertheless, the SDR formulas (7) and (8) give good approximations of $R_2$ and $r_2$ as long as $5 < \Delta\omega\tau_D < 20$ and a provide good estimate of the maximum values reached in the middle of the SDR range, when $\Delta\omega\tau_D \approx 10$.[26] Above an upper limit $\Delta\omega\tau_D \approx 20$, $R_2$ is no more approximated by the SDR, since the refocusing pulses used in the $T_2$ measurement sequence become effective. In this third relaxation regime described by the Partial Refocusing Model (PRM), $R_2$ is lower than $R_2^*$ and exhibits a strong dependence on the echo time chosen for the measuring sequence.[28]

For particles with characteristics corresponding to $1 < \Delta\omega\tau_D < 5$, neither **Equation (3)** nor Equation (8) is valid and a transition between MAR and SDR relaxation is observed. As $\Delta\omega\tau_D > 20$, one observes a transition between SDR and PRM. **Figure 2** represents the normalized relaxivities $r_2' = r_2\times\phi_{intra}$ for all the systems with $\Delta\omega\tau_D > 1$ (see Table 1). Only particles of approximately same $M_v$ can be compared together. The figure also gives – for different $M_v$ ranges – the predictions of the empirical function recently validated by computer simulations for particles outside the MAR.[26] The trend of these empirical curves is compatible with the experimental data. It should be stressed that the comparison between all the samples – and also with theory – is difficult since these systems are in a "theoretical no-man's-land" and also because, for many samples, $R_2$ surely depends on the echo-time of the measurement sequence. Nevertheless, most experimental data are located in the domains of Figure 2 corresponding to their size and magnetization, except those derived from the article by C. Paquet _et al_ on magnetic hydrogels,[15] which show higher $r_2\times\phi_{intra}$ values than expected from the simulations (for particle sizes ~150 nm and 175 nm). The simulations shown here do not take into account the mechanism reported recently for highly hydrated particles.[27] In that case, the slowing down of proton diffusion near the magnetic particle surface – where the magnetic gradients are the strongest – induces fast proton dephasing, which significantly raises the relaxivity.[27] The present study holds for the two limit cases of either impermeable coatings or completely permeable shells. Therefore "smart" coatings modifying proton diffusion such as the magnetic hydrogels somehow deviate from the general behavior.

### 2.3.3. _Maximal achievable transverse relaxivities_

As learned from computer simulations[26] and supported by the data presented in Figure 2, the relaxation enhancement effect of magnetic particles is expected to reach a maximum value for a particular size. This maximum occurs when the system is completely in the SDR. As previously stated, this is the case when $\Delta\omega\tau_D \approx 10$. It is thus possible to estimate, using Equation (8), the maximum efficiency of different types of magnetic particles as well as the optimal size at which this maximum should be reached. The results, obtained by using the expressions of $\Delta\omega$ and $\tau_D$, are presented in **Table 2**.

The first important information obtained from Table 2 is that the maximum relaxivity is the same for maghemite nanoparticles and clusters of maghemite particles, only the optimal size of particles needed to reach the maximum is different (obtained through the condition $\Delta\omega\tau_D \approx 10$). In the case of individual magnetic cores, such a diameter of 55 nm falls above the classical limit size of 40 nm for USPIOs defined relatively to the sizes of biological barriers, but remains below the maximum size of magnetic (Weiss) monodomains.[25] Moreover, it will be a challenge to reach it experimentally in a proper dispersed (colloidal) state because of strong inter-particular attractions. In the literature, different clusters of $\gamma$–$Fe_2O_3$ or $Fe_3O_4$ cores[15-18] plotted on Figure 2 already exceeded a $T_2$ relaxivity of 500 $s^{-1}$·$mM^{-1}$ but remained below the theoretical maximum of 750 $s^{-1}$·$mM^{-1}$ not yet reached experimentally, according to the authors' knowledge. Other studies presenting dense magnetite clusters around 100 nm only show a moderate increase of $r_2$ compared to the individual USPIOs,[29] presumably due to too high size dispersity. Secondly and as expected,[22,30] core-shell or special compositions of ferrite particles with a higher saturation magnetization than pure iron oxide led to higher relaxivities. It was recently proven that an appropriate composition of zinc and manganese ferrite enables reaching $r_2$=860 $s^{-1}$·$mM^{-1}$ for a diameter $d$=15 nm.[30] According to Table 2, $r_2$ should reach even larger values (up to 1200–1860 $s^{-1}$·$mM^{-1}$) by increasing the size both for $Fe@Fe_2O_3$ core-shells and $(Zn_{0.4}Mn_{0.6})Fe_2O_4$ mixed ferrite nanoparticles. Increasing the specific magnetization of USPIOs by an appropriate choice of their metal composition is indeed an interesting option to optimize MRI contrast agents, as long as toxicity is not introduced by the non ferrous metals.

## 3. Conclusions

Despite the broad variety of superparamagnetic MRI contrast agents differing by their size, geometry (filled micelles or hollow vesicles, dense or loose clusters…), type of coating (organic or inorganic, impermeable or porous, hydrophilic or hydrophobic…), no specific models need to be introduced. We have indeed evidenced in this article that the classical MAR and SDR models can correctly represent the experimental data once structural and magnetic parameters are known (external diameter, volume fraction and magnetization of the magnetic materials) and the relaxivity is appropriately normalized. More precisely, the MAR is verified by individual USPIOs or clusters which are either compact or diluted in a non magnetic material. In the latter case, the porosity (*e.g.* silica) or permeability to water (*e.g.* hydrogel) is not an issue: such internal protons relax much faster than external ones, but their contribution to the measured relaxation rate remains limited due to their low volume fraction compared to the water protons diffusing in the "outer shell" around the particle. The relaxivity at high magnetic field / Larmor frequency of particles in the MAR follows a universal scaling law varying with the square of diameter, square of magnetization and inverse of the internal volume fraction of magnetic material. The experimental prefactor of this power law is in good accordance with the physical constants of the models. For larger or more concentrated clusters, the SDR model correctly describes the plateau value that is observed experimentally. Moreover, the size and magnetization of the particle can be chosen to satisfy the condition $\Delta\omega\tau_D \approx 10$ in order to design contrast agents of maximum $T_2$ relaxivity.[26] But it should be stressed that this optimum $r_2$ will only be approached by particles presenting a rather narrow size distribution centered on the optimal size, since smaller (in motional averaging regime) and larger particles (following the Partial Refocusing Model) will present lower efficiencies and decrease the mean transverse relaxivity of the sample. This explains why some already reported SPIO clusters did not exhibit tremendous $r_2$ despite mean values of magnetization, volume fraction and hydrodynamic diameter close to the optimal ones.[29]

To conclude, by validating simple principles of the theory of proton relaxation on a wide range of experimental systems, this article proposes a unified method to predict the transverse relaxivity $r_2$ of MRI contrast agents at clinical field based on materials ($M_v$) and geometrical ($d$, $\phi_{intra}$) parameters. These results offer practical tools to the chemists who aim at optimizing the relaxation properties for MRI in the design of more elaborated particles than the commercially available $T_2$ contrast agents, such as multi-modal probes or theranostic nanovectors.

## 4. Experimental

*Synthesis of Ultra-small Superparamagnetic Iron Oxide.* USPIO nanoparticles made of maghemite (γ-$Fe_2O_3$) were synthesized in water according to Massart's procedure.[2] At first, magnetite $Fe_3O_4$ nanocrystals (also called ferrous ferrite $FeO.Fe_2O_3$) were prepared from an alkaline coprecipitation of a quasi-stoichiometric mixture of iron +II (0.9 mol) and iron +III (1.5 mol) chloride salts in HCl solution (3L, pH≈0.4). One liter of a concentrated ammonia solution (7 mol) was quickly added onto the acidic iron salts mixture, which produced a black solid suspension almost instantaneously. After 30 minutes of stirring at 800 rpm, the $Fe_3O_4$ nanoparticles were attracted by a strong ferrite magnet (152×101×25.4 $mm^3$, Calamit Magneti, Milano-Barcelona-Paris). Then the supernatant (≈2.25 L) containing non magnetic ferrihydrites (reddish flakes) was discarded and the magnetic precipitate (black) was washed with 1 L water. After sedimentation on the ferrite magnet, the flocculate was acidified with 0.26 L of nitric acid (69%) and stirred 30 min after being completed up to 2 L with water. In order to be completely oxidized from magnetite into maghemite, the solid phase was separated from the supernatant (≈1.5 L, red) and immersed in a boiling solution of ferric nitrate (0.8 mol in 0.8 L). After 30 min under stirring at 90-100 °C, the suspension had turned into the red colour characteristic of maghemite γ-$Fe_2O_3$. After washing steps in acetone and diethyl-ether to remove the excess ions, the nanoparticles readily dispersed in water and formed a true "ionic ferrofluid" made of maghemite nanoparticles. The iron oxide surface bore positive charges due to adsorption of protons in acidic media, in that case a dilute $HNO_3$ solution at pH between 1.2 and 1.7. Therefore such ferrofluid remains in a monophasic state under the application of a magnetic field of arbitrary value.[2–5]

*Size sorting.* On the microscopic scale, those crystals exhibit a Log-Normal distribution of diameters of parameters $d_0$ =7 nm and $\sigma$ =0.38, as measured by magnetometry.[3] Maghemite nanoparticles with such a high size-dispersity can be treated with a size-sorting procedure based on fractionated phase-separation.[4] More precisely, the addition of an excess of $HNO_3$ not only lowers the pH but also raises the ionic strength, thereby screening the electrostatic repulsions between the nanoparticles. Above a threshold electrolyte concentration, a liquid-liquid phase separation occurs between a concentrated "liquid-like" phase and a dilute "gas-like" phase. After magnetic sedimentation on a strong ferrite magnet (152×101×25.4 $mm^3$, Calamit Magneti, Milano-Barcelona-Paris) to accelerate demixtion, a concentrate (denoted C1) could be readily separated from the supernatant (denoted S1). Once washed with acetone to remove the excess of ions, the two separated fractions were dispersed in water. The fit of their magnetization curve by VSM leads to their size distributions modelled by a Log-normal law with $d_0$ as median diameter and $\sigma$ as standard width of the logarithms of diameters: $d_0$ = 8.7 nm ($\sigma$ =0.35) for C1 and $d_0$ = 7.1 nm ($\sigma$ =0.29) for S1. The enrichment of the "liquid-like" phase by the larger size tail of the distribution compared to the dilute "gas-like" phase originates from the sensitivity of the inter-nanoparticle potential with the diameters (the larger nanoparticles exhibiting much higher Van der Waals interactions between them). By repeating the phase-separation protocol on both samples C1 and S1, we obtained four new fractions at second level of refined distribution of sizes, and so on after a third and fourth level as indicated on **Sketch S1**, among which several fractions were used in the following of the article either as they were in nitric acid conditions (C1S2, C1C2C3, C1C2S3, S1S2C3) or after coating with citrate or polyacrylate ligands (S1S2S3, S1S2C3, C1C2C3C4). Unlike the preceding steps for which fractions were divided into culots (C) and supernatants (S) by phase separation under increased $HNO_3$ concentration, for the final step C1C2C3C4S5 another method was used. Namely S5 stands in that case for "sedimentation": a strong magnetic gradient was used indeed to induce a vertical concentration gradient (but not a true separation into 2 phases as with the electrolyte). A colloidal suspension enriched in the largest magnetic nanoparticles was pipetted at the bottom of the cuvette.[31] After the size-sorting process, the fractions of interest (S1S2S3, S1S2C3, C1C2C3C4 and C1C2C3S4) were coated either with tri-sodium citrate[5] $Na_3Cit$ or sodium polyacrylate[6,32] ($PAA_{2k}$ or $PAA_{5k}$) using a protocol based on electrostatic complexation and adsorption.

*Coating with citric acid.* Briefly, the grafting was made by reacting 37.4 g of $Na_3Cit$ per mole of iron oxide (20% molar) around pH 8 for 30 min at 70 °C under vigorous stirring and subsequent removal of the supernatant by magnetic sedimentation. Then three washing cycles were performed with acetone to remove the excess ions and finally with diethyl ether to remove acetone. The obtained precipitate of citrate-coated USPIOs can be readily suspended in pure water by simple vortexing. To insure a perfect colloidal stability with low hydrodynamic diameters as probed by dynamic light scattering, the salinity caused by unbound citrate ions was decreased by dialysis for 24 hours against 8 mM $Na_3Cit$.

*Coating with Poly(acrylic acid).* Poly(sodium acrylate), the salt form of polyacrylic acid, with a molar mass $M_n = 2000$ g mol$^{-1}$ ($PAA_{2k}$) or $M_n = 5000$ g mol$^{-1}$ ($PAA_{5k}$) and a polydispersity index $M_w/M_n = 1.7$ was purchased from Sigma Aldrich (references 81130 and 81132) and used without further purification. In order to adsorb polyelectrolytes onto the surface of the nanoparticles, we followed the "precipitation-redispersion" protocol.[6,32] The precipitation of the cationic iron oxide dispersion by $PAA_{2k/5k}$ was performed in acidic conditions (pH 2) at weight concentrations of 1 g·L$^{-1}$ for both nanoparticles and polymer. The precipitate was separated from the solution by centrifugation, and its pH was increased by addition of ammonium hydroxide. The precipitate redispersed spontaneously at pH≈7–8, yielding a clear solution that now contained the polymer-coated particles. The hydrodynamic sizes of $\gamma$-Fe$_2$O$_3$ USPIOs coated by $PAA_{2k}$ were found to be 5 nm larger than the hydrodynamic diameter of the uncoated particles, indicating a corona thickness $h = 2.5$ nm.[32] In terms of coverage, the number of adsorbed chains per particle was estimated to be 1 nm$^{-2}$ (assuming a 1:1 $PAA_{2k}$-iron oxide weight ratio for an USPIO of molar mass ≈$10^6$ g·mol$^{-1}$ and surface ≈500 nm$^2$). As a final step, the dispersions were dialyzed against DI-water which pH was first adjusted to 8 (Spectra/Por 2 dialysis membrane with MWCO 12 kD). At this pH, 90 % of the carboxylate groups of the PAA coating were ionized. Electrophoretic mobilities were found at values $\mu_E = -3.76\times10^{-4}$ and $-3.52\times10^{-4}$ cm$^2$ V$^{-1}$ for Cit–$\gamma$-Fe$_2$O$_3$ and $PAA_{2k}$–$\gamma$-Fe$_2$O$_3$ respectively. As a final step of the procedures described above, the dispersions were dialyzed against DI-water which pH was first adjusted to 8 by addition of sodium hydroxide (Spectra Por 2 dialysis membrane with MWCO 12 kD). For the citrate-coated particles, DI-water was supplemented with 8 mM of free citrates. At this pH, 90 % of the carboxylate groups of the citrate and $PAA_{2K}$ coating were ionized. The suspension pH was adjusted with reagent-grade nitric acid ($HNO_3$) and with sodium or ammonium hydroxides. For the assessment of the stability with respect to ionic strength ($I_S$), sodium and ammonium chloride (NaCl and NH$_4$Cl, Fluka) were used to control $I_S$ in the range 0 – 1 M.[32]

*Clustering.* Different clusters were prepared with the S1S2C3 iron oxide cores. The principle consists in mixing negatively charged USPIOs coated with PAA and a double-hydrophilic diblock copolymer (DHBC) such as poly(trimethylammonium ethylacrylate methylsulfate)-*b*-polyacrylamide (PAM$_{30k}$-*b*-PTEA$_{11k}$) made of a neutral block (PAM) and a cationic block (PTEA) at a high salt concentration where the electrostatic interactions are totally screened. When the salinity is decreased at a controlled rate either by dilution or dialysis, below a threshold concentration ([NH$_4$Cl]=0.4 mol/L) a microphase separation occurs by association between the oppositely charged species. The coacervates are perfectly spherical, with a magnetic core containing a limited number of USPIOs wrapped by a neutral polymer shell that prevents further aggregation. Different clusters of varying size and magnetization can be prepared by varying the salinity decrease rate.[32] For example, spherical SPIO particles containing approximately 70 iron oxide crystals with a hydrodynamic diameter of 127 nm (see also TEM picture on **Figure S2**) were obtained by mixing S1S2C3@PAA$_{2k}$ USPIOs (0.75 g·L$^{-1}$) and PAM$_{30k}$-*b*-PTEA$_{11k}$ (1.5 g·L$^{-1}$) in 0.43 mol·L$^{-1}$ NH$_4$Cl and diluting 3 times with pure water (down to 0.143 mol·L$^{-1}$ NH$_4$Cl). Iron oxide represents 33 % w/w of these hybrid particles and thus an average volume fraction of 6 % v/v only.

*Vibrating sample magnetometry.* A laboratory made VSM instrument was used, measuring the magnetization curve *versus* excitation *M(H)* at RT for a magnetic suspension of volume fraction *f* from the signal induced in detection coils when the sample is moved periodically in an applied magnetic field varied from 0 to 1 T (thanks to synchronous detection and with an appropriate calibration to convert the signal in mV into A·m$^{-1}$).

*Dynamic light scattering.* DLS measurements were performed on a Malvern NanoZS apparatus operating at a 173 ° scattering angle. The collective diffusion coefficient *D* was determined from the second-order autocorrelation function of the scattered light. From the value of the coefficient, the hydrodynamic diameter of the colloids was calculated according to the Stokes-Einstein relation, $d_H = k_B T / 3\pi\eta_S D$, where $k_B$ is the Boltzmann constant, *T* the temperature (*T* = 298 K) and $\eta_S$ the solvent viscosity ($\eta_S = 0.89 \times 10^{-3}$ Pa s for water). The autocorrelation functions were interpreted using the 2$^{nd}$ order cumulants (Z-average diameter and Poly-Dispersity Index) and the multimodal fit provided by the instrument software.

*Relaxometry.* Relaxation time measurements were performed at low fields on BRUKER (Germany) mq 20, mq 60 instruments and a Spintrack relaxometer operating at magnetic fields (B$_0$) of 0.47, 1.41, 0.67 and 0.93 T respectively. BRUKER AVANCE-200 (4.7 T), BRUKER AMX 300 (7 T) and AMX 500 (11.7 T) spectrometers were used for the high-field $T_1$ measurements. $T_1$ relaxation profiles were recorded at 5 °C and 37 °C from 0.00023 to 0.23 T on a Spinmaster fast field cycling relaxometer (STELAR, Mede, Italy). In most of the graphs, the magnetic field is expressed in term of proton Larmor frequency: a field of 1 Tesla corresponds to a Larmor frequency of 42.6 MHz. The results are expressed as longitudinal and transverse relaxivities which are defined by the increase in the longitudinal and transverse relaxation rates due to an increase of 1 mM in the paramagnetic ion concentration. The relaxivity values were properly calculated by an accurate titration of the iron content in all samples as determined using Inductively Coupled Plasma Atomic Emission spectroscopy (Thermo, USA), after micro-wave mineralization of the suspensions with a mixture of nitric acid and hydrogen peroxide. Varying shapes of the $T_1$ profiles reflect the differences in sizes, magnetizations and Néel relaxation times of the particles.[33] $T_2$ was measured with a CPMG sequence using an inter-echo time of 1 ms. $T_2$* was not measured for our samples. Indeed, different tests on the high resolution spectrometers showed that even for reference solutions (simply containing gadolinium ions, for example), the value of $T_2$* evaluated with the line-width of the resonance peak was always significantly lower than the $T_2$ value, while for such systems $T_2$ and $T_2$* should be identical. The influence of the shims seems to be critical when estimating $T_2$* with this technique. For strongly magnetic compounds, the measurement of a "real" $T_2$*, comparable to the value predicted by the different microscopic relaxation theories, could be very difficult.

*Numerical simulation.* The empirical expression used to plot the solid curves on Figure 2 was validated by a previously described Monte Carlo simulation of $T_2$ decay at high fields.[26] The methodology consists mainly in three steps. Firstly, static and impenetrable spherical magnetic particles are distributed in the simulation space. Secondly, the diffusion of each proton is simulated by a random walk. At each time step, the spin dephasing of each proton – proportional to local dipolar magnetic field produced by the particles – is computed. Finally, the MR signal decay is obtained by averaging all the protons spins and an exponential law can be fitted to the data to obtain the transverse relaxation rate $r_2$.

**Supporting Information**

Supporting Information is available online from the Wiley Online Library or from the author. S1. Synoptic scheme of the size sorting procedure. S2. Characterization of the nanoparticles: a) Magnetometry; b) Transmission Electron Microscopy; c) NMR Relaxometry.

**Acknowledgements**

Author-One and Author-Four are grateful to Dr Alain Roch for helpful discussions and to Prof. Dr. Robert N Muller for the access to the 60 MHz relaxometer. Authors also thank Aude Michel and Delphine Talbot (PECSA) respectively for TEM and iron atomic emission spectroscopy. This research was supported in part (UMONS) by the Fonds de la Recherche Scientifique (IISN 4.4507.10), (MSC/PECSA/LCPO) by the Agence Nationale de la Recherche under the contracts BLAN07-3_206866 "ITC-nanoProbe", and (MSC/PECSA) by the European Community through the project "NANO3T—Biofunctionalized Metal and Magnetic Nanoparticles for Targeted Tumor Therapy", project number 214137 (FP7-NMP-2007-SMALL-1). Author-One is a research fellow from FRS-FNRS.

**Annotations**

CA, contrast agents; CPMG, Carr-Purcell-Meiboom-Gill; DLS, dynamic light scattering; MAR, motional averaging regime; MRI, magnetic resonance imaging; NMRD, nuclear magnetic resonance dispersion; OS, outer sphere; SDR, static dephasing regime; SI, système international d'unités; SPION, superparamagnetic iron oxide nanoparticle; USPIO, ultra-small superparamagnetic iron oxide; VSM, vibrating sample magnetometry.

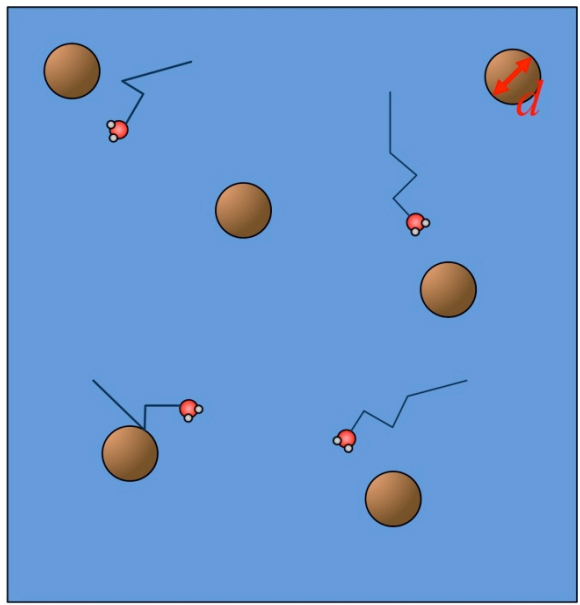

(a)

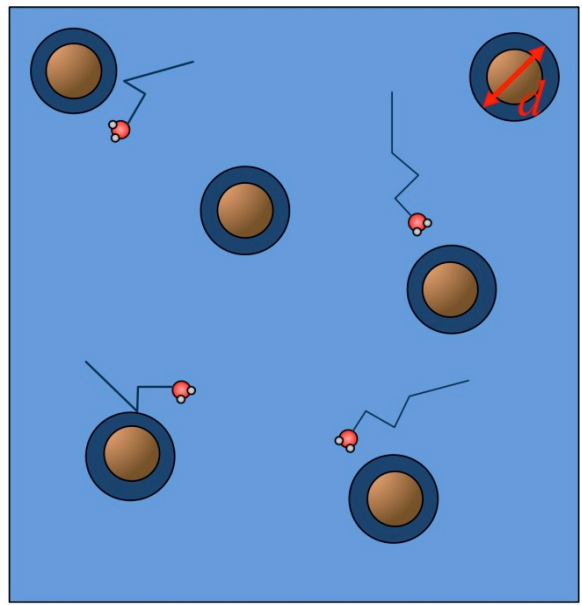

(b)

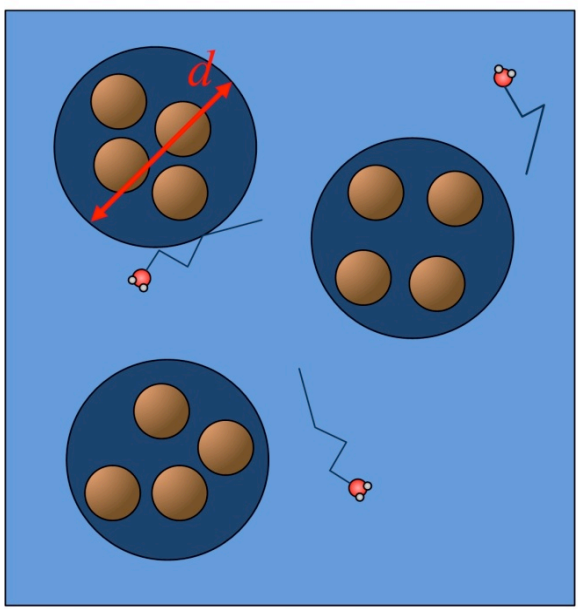

(c)

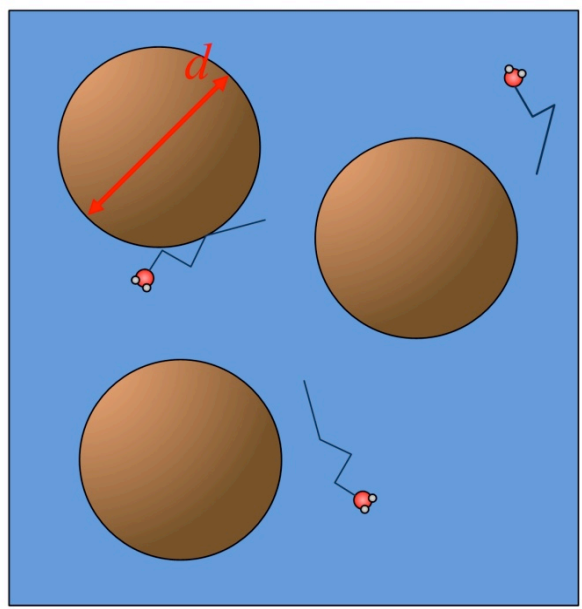

(d)

**Scheme 1.** In the commonly used relaxation models, $d$ is defined as twice the minimum approach distance of a water molecule to the center of the contrast agent. For a single nanoparticle, $d$ is equal to its diameter (case a). If there is a layer inaccessible to water molecules (case b), $d$ must include this impermeable coating thickness (dark blue shell). If nanoparticles are clustered (case c), $d$ corresponds to the whole hydrodynamic diameter, determined by dynamic light scattering. The MAR model can predict the relaxivities induced by such systems, assimilating the cluster to a single magnetic particle with adapted diameter $d$ and magnetization $M_v$ (case d). But in order to compare with MAR equations, the $r_2$ relaxivity has to be multiplied by the intra-aggregate volume fraction $\phi_{intra}$ to get a corrected relaxivity $r_2$' induced by the aggregate (see text).

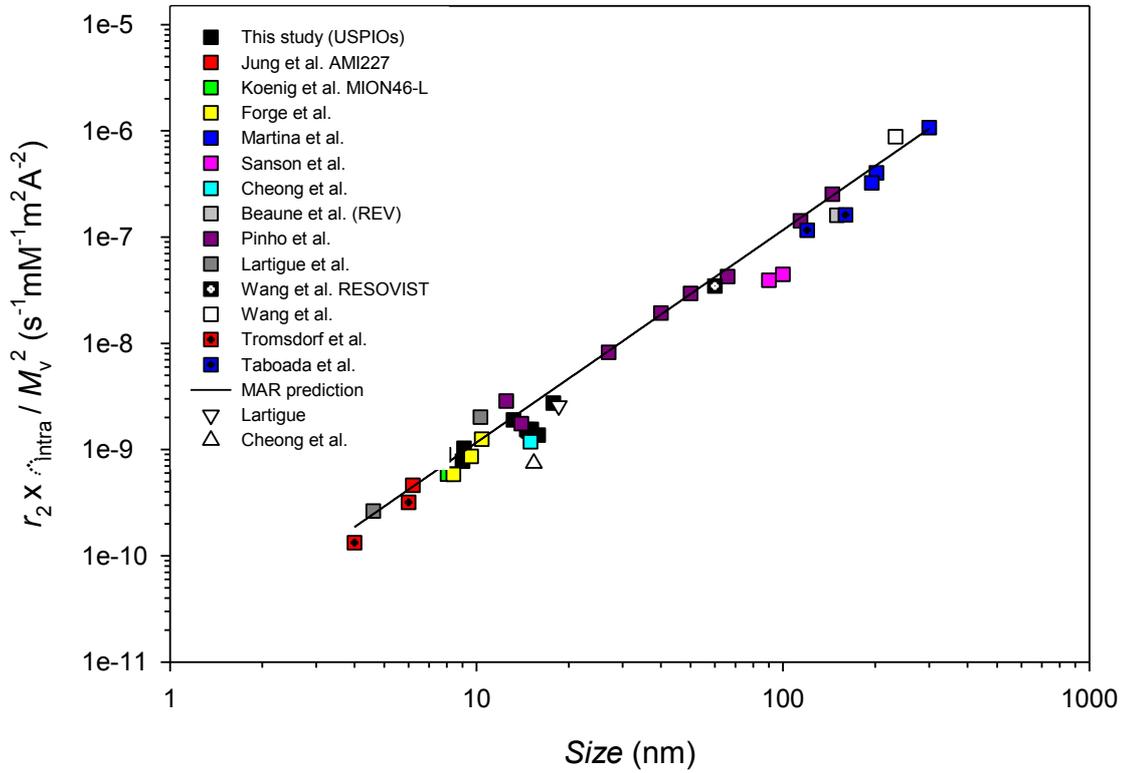

**Figure 1.** Samples in the motional averaging regime (MAR), with $\Delta\omega\tau_D < 1$. Influence of the size on transverse relaxivity at high field ($\geq$1T) and at 37°C with the appropriate weighting by the intra-aggregate volume fraction $\phi_{Intra}$ and normalization by the square of the magnetization enabling to compare USPIOs (single cores), maghemite core-silica shell particles, magnetic vesicles and SPIOs (clusters) on the same curve. The solid line is Equation (6).

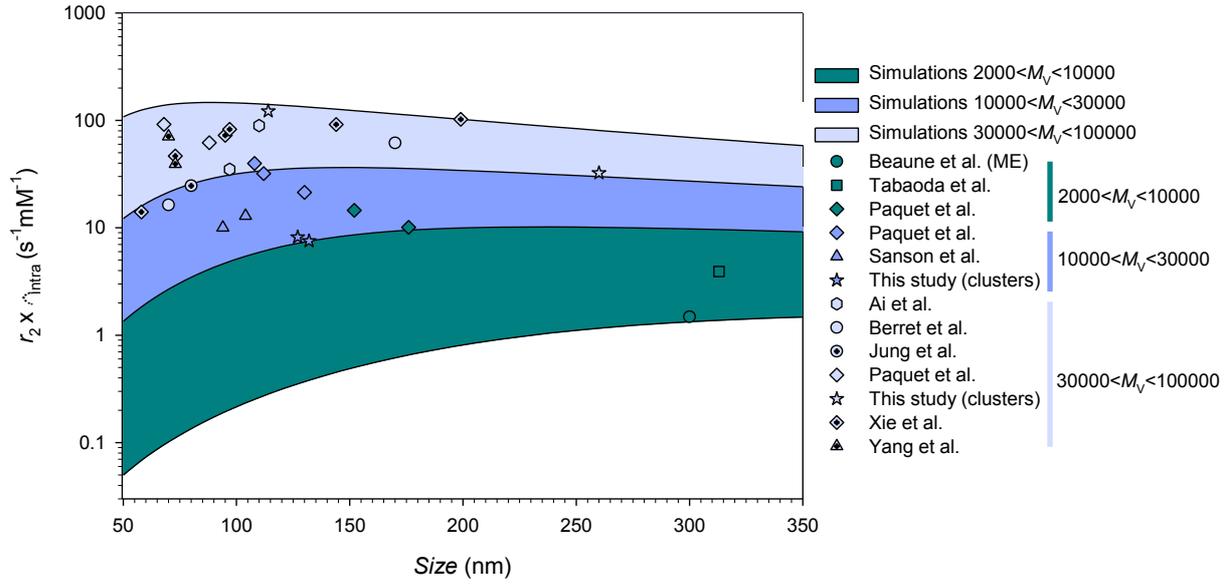

**Figure 2.** Samples out of the MAR, with $\Delta\omega\tau_D > 1$. Influence of the size on transverse relaxivity at high field ($\geq$1T) and at 37°C with the appropriate weighting by the intra-aggregate volume fraction $\phi_{intra}$. The colored regions correspond to the relaxivities obtained by computer simulations, through the use of an empirical function which was recently validated.[26]

**Table 1** summarizes the parameters used for the samples presented in **Figure 1** and **Figure 2**.

| Reference | Sample code | Magnetic materials | Nature of coating | Size (nm) | Method | $\phi_{intra}$ | $M_v$ (A·m⁻¹) | $r_2$ (s⁻¹·mM⁻¹) | $\phi_{intra} \cdot r_2 \cdot M_v^{-2}$ (s⁻¹·mM⁻¹·A⁻²·m²) | $\Delta\omega\tau_D$ |
|---|---|---|---|---|---|---|---|---|---|---|
| **In motional averaging regime – $\Delta\omega\tau_D<1$** | | | | | | | | | | |
| This study | S1S2S-PAA2k | γ-Fe₂O₃ | Hydrophilic Polymer (PAA2k or PAA5k) | 7.8 | VSM-d$_w$ | 1 | 280000 | 70 | $8.9 \cdot 10^{-10}$ | 0.16 |
| | C1C2-PAA2k | | | 14.6 | | | 370000 | 211.46 | $1.54 \cdot 10^{-9}$ | 0.74 |
| | S1S2C3-PAA2k | | | 9.1 | | | 290000 | 86.5 | $1.03 \cdot 10^{-9}$ | 0.22 |
| | C1C2C3C4S5-PAA5k | | | 17.8 | | | 327000 | 292.6 | $2.74 \cdot 10^{-9}$ | 0.97 |
| | S1-PAA2k | | | 9 | | | 300000 | 69.8 | $7.8 \cdot 10^{-10}$ | 0.23 |
| | C1C2C3 | | adsorbed H⁺ only | 15.1 | | | 342000 | 181.6 | $1.55 \cdot 10^{-9}$ | 0.73 |
| | C1S2S3 | | | 14.8 | | | 333000 | 166 | $1.50 \cdot 10^{-9}$ | 0.68 |
| | C1S2 | | | 13.2 | | | 328000 | 205 | $1.90 \cdot 10^{-9}$ | 0.53 |
| | C1C2C3C4-cit | | Citrate | 15.9 | | | 312000 | 133.3 | $1.37 \cdot 10^{-9}$ | 0.73 |
| Jung et al [13] | AMI227 Ferumoxtran Sinerem (brand names) | γ-Fe₂O₃ /Fe₃O₄ mixture | Hydrophilic Polymer (Dextran) | 6.2 | TEM-d$_w$ | 1 | 339000 | 53 | $4.6 \cdot 10^{-10}$ | 0.13 |
| Koenig et al [9] | MION46L Clariscan (brand names) | | | 8.05 | VSM | 1 | 274000 | 44 | $5.9 \cdot 10^{-10}$ | 0.17 |
| Forge et al [8] | 250 Gauss | γ-Fe₂O₃ | aminoproyl-trimethoxy-silane | 10.4 | VSM | 1 | 326000 | 133 | $1.24 \cdot 10^{-9}$ | 0.33 |
| | 1000 Gauss | | | 9.6 | | | 332000 | 95.2 | $8.64 \cdot 10^{-10}$ | 0.29 |
| | 2000 Gauss | | | 8.4 | | | 306000 | 54.7 | $5.84 \cdot 10^{-10}$ | 0.20 |
| Lartigue et al [10] | Rha-4 | γ-Fe₂O₃ | rhamnose-phosphonate | 4.6 | NMRD | 1 | 399000 | 42 | $2.64 \cdot 10^{-10}$ | 0.08 |
| | Rha-10 | | | 10.3 | | | 363000 | 266 | $2.02 \cdot 10^{-9}$ | 0.36 |
| | Rha-18 | | | 18.5 | | | 337000 | 292 | $2.57 \cdot 10^{-9}$ | 1.1 |
| Tromsdorf et al [24] | 4nm-PEG1100 | Fe₃O₄ | PEG-phosphate | 4 | TEM | 1 | 363000 | 17.5 | $1.3 \cdot 10^{-10}$ | 0.05 |
| | 6nm-PEG1100 | | | 6 | | | | 42 | $3.2 \cdot 10^{-10}$ | 0.12 |
| Martina et al [19] | Conventional MFL | γ-Fe₂O₃ | Citrate-coated Magnetic Fluid encapsulated in Liposomes | 300 | DLS | $5 \cdot 10^{-4}$ [a] | 177 | 67 | $1.08 \cdot 10^{-6}$ | 0.15 |
| | PEG-ylated MFL | | | 202 | | $2.6 \cdot 10^{-3}$ [a] | 896 | 124 | $4.06 \cdot 10^{-7}$ | 0.34 |
| | | | | 195 | | $3.3 \cdot 10^{-3}$ [a] | 1150 | 130 | $3.24 \cdot 10^{-7}$ | 0.41 |
| Sanson et al [21] | WDi-20 | γ-Fe₂O₃ | Hydrophobic Membrane of Polymersome (PTMC-PGA) | 100 | DLS | $2.3 \cdot 10^{-2}$ [b] | 6460 | 81 | $4.49 \cdot 10^{-8}$ | 0.60 |
| | WDi-35 | | | 90 | | $4.4 \cdot 10^{-2}$ [b] | 12200 | 134 | $3.93 \cdot 10^{-8}$ | 0.92 |
| Cheong et al [22] | oxide | γ-Fe₂O₃ | Dimercapto-succinic acid (DMSA) | 15 | TEM | 1 | 350000 | 145 | $1.2 \cdot 10^{-9}$ | 0.74 |
| Beaune et al [20] | DDAB magnetic vesicles: REV | γ-Fe₂O₃ | Oleic acid (OA) | 150 | TEM | $1.6 \cdot 10^{-2}$ [c] | 4200 | 177 | $1.6 \cdot 10^{-7}$ | 0.87 |

[a] From the iron/lipid molar ratio converted into a weight ratio inside the membrane. Assuming a lipid density of 1 and a membrane thickness of 3.5 nm, we deduced the volume of iron oxide inside the whole volume of the vesicle.
[b] From the feed weight ratios inside the membrane (20%, 35%, 50 % and 70%). Assuming a polymer density of 1 and a membrane thickness of 10 nm (SANS), we deduced the volume of iron oxide inside the whole volume of the vesicle.
[c] Measured by magnetophoresis. REV stands for "reverse phase evaporation", ME for "multiple emulsion", DDAB for didodecyldimethyl ammonium bromide. The USPIOs have a $d_w$=7.5 nm, $M_v$=2.6·10⁵ A/m and $r_2$=105 s⁻¹m·M⁻¹.

| Reference | Sample code | Magnetic materials | Nature of coating | Size (nm) | Method | $\phi_{intra}$ | $M_v$ (A·m⁻¹) | $r_2$ (s⁻¹·mM⁻¹) | $\phi_{intra}·r_2·M_v^{-2}$ (s⁻¹·mM⁻¹·A⁻²·m²) | $\Delta\omega\tau_D$ |
|---|---|---|---|---|---|---|---|---|---|---|
| **In motional averaging regime – $\Delta\omega\tau_D<1$ (continued)** | | | | | | | | | | |
| Pinho et al [11] | FF (core) | γ-Fe₂O₃ | Porous SiO₂ | 12.5 | VSM [d] | 1 | 282000 | 228 | 2.11·10⁻⁹ | 0.41 |
| | 0A | | | 14 | TEM | 0.71 | 201000 | 100 | 1.29·10⁻⁹ | 0.37 |
| | 1A | | | 27 | | 0.1 | 27900 | 64 | 6.37·10⁻⁹ | 0.19 |
| | 2A | | | 40 | | 3·10⁻² | 8550 | 47 | 1.44·10⁸ | 0.13 |
| | 3A | | | 50 | | 1.5·10⁻² | 4400 | 38 | 2.24·10⁻⁸ | 0.10 |
| | 4A | | | 66 | | 6.8·10⁻³ | 1920 | 23 | 3.12·10⁻⁸ | 0.08 |
| | 5A | | | 114 | | 1.3·10⁻³ | 370 | 15 | 1.05·10⁻⁷ | 0.04 |
| | 6A | | | 145 | | 6.1·10⁻⁴ | 170 | 12 | 1.95·10⁻⁷ | 0.03 |
| Wang et al. [17] | IOs (Resovist) | γ-Fe₂O₃ | Hydrophilic Polymer (Dextran) | 60 | DLS | 8.4·10⁻² [e] | 19500 | 282.4 | 3.5·10⁻⁸ | 0.65 |
| | IO-loaded PLGA-mPEG | | Amphiphilic Copolymer Micelle (PLGA-PEG) | 233 | | 4·10⁻³ [e] | 1160 | 532.7 | 8.8·10⁻⁷ | 0.59 |
| Taboada et al. [12] | S1 | γ-Fe₂O₃ | Porous SiO₂ (aerogel) | 160 | DLS | 1.1·10⁻² | 3180 | 148 | 1.6·10⁻⁷ | 0.76 |
| | S2 | | | 120 | | 1.5·10⁻² | 4590 | 164 | 1.16·10⁻⁷ | 0.62 |
| **Out of motional averaging regime – $\Delta\omega\tau_D>1$** | | | | | | | | | | |
| Taboada et al. [12] | S3 | γ-Fe₂O₃ | Porous SiO₂ (aerogel) | 313 | DLS | 1.2·10⁻² | 3880 | 326 | 2.6·10⁻⁷ | 3.6 |
| This study | Clusters S1S2C | γ-Fe₂O₃ cluster | Double Hydrophilic Copolymer (PAM-PTEA) | 127 | DLS | 9·10⁻² | 26000 | 91 | 1.22·10⁻⁸ | 3.9 |
| Beaune et al [20] | DDAB magnetic vesicles: ME | γ-Fe₂O₃ | Oleic acid (OA) | 300 | TEM | 8·10⁻³ [c] | 2100 | 185 | 3.36·10⁻⁷ | 1.6 |
| This study | S2C14 | γ-Fe₂O₃ cluster | Double Hydrophilic Copolymer (PAM-PTEA) | 132 | DLS | 8.3·10⁻² | 24200 | 91 | 1.29·10⁻⁸ | 3.9 |
| | S2C12 | | | 260 | | 0.15 | 44000 | 216 | 1.7·10⁻⁸ | 28 |
| | C1C2C3C4-PAA5k | γ-Fe₂O₃ demixted droplet | Hydrophilic Polymer (PAA) | 114 | | 0.286 | 89000 | 427 | 1.54·10⁻⁸ | 10.8 |
| Berret et al [14] | PTEA(5k) | γ-Fe₂O₃ cluster | Double Hydrophilic Block Copolymer (PAM-PTEA) | 70 | DLS | 0.22 | 57200 | 74 | 4.98·10⁻⁹ | 2.2 |
| | PTEA(11k) | | | 170 | | 0.38 | 99000 | 162 | 6.28·10⁻⁹ | 25 |
| Xie et al [18] | Loosen SPION cluster (A) | Fe₃O₄ cluster | Oleic acid (OA) / Oleylamine in Amphiphilic Copolymer Micelle (PEG-PLA) | 58 | DLS | 0.12 [f] | 42000 | 117 | 8·10⁻⁹ | 1.3 |
| | Condense cluster B | | | 73 | | 0.20 [f] | 70000 | 233 | 9.5·10⁻⁹ | 3.5 |
| | Condense cluster C | | | 95 | | | | 363 | 1.48·10⁻⁸ | 5.9 |
| | Condense cluster D | | | 97 | | | | 413 | 1.69·10⁻⁸ | 6 |
| | Condense cluster E | | | 144 | | | | 458 | 1.87·10⁻⁸ | 14 |
| | Condense cluster F | | | 199 | | | | 512 | 2.09·10⁻⁸ | 26 |

[d] We are in debt to the authors of this reference to have provided us their ferrofluid to make this VSM measurement. Volume fractions were calculated from TEM images. Relaxivities at 25 °C were rescaled by a factor 0.686/0.895.

[e] Calculated from the iron weight ratio measured by inductively coupled plasma mass spectroscopy (ICP-MS). The volume fractions enable to get the volume magnetizations using given values of specific magnetizations (72.9 emu·g⁻¹ for IOs and 83.5 emu·g⁻¹ for IO-loaded PLGA-mPEG).

[f] Measured by thermogravimetry analysis (TGA). The USPIOs have a $d_w$=10.5 nm (TEM, $N$=1822) and $M_v$=3.3·10⁵ A·m⁻¹.

| Reference | Sample code | Magnetic materials | Nature of coating | Size (nm) | Method | $\phi_{Intra}$ | $M_v$ (A·m⁻¹) | $r_2$ (s⁻¹·mM⁻¹) | $\phi_{Intra}\cdot r_2 \cdot M_v^{-2}$ (s⁻¹·mM⁻¹·A⁻²·m²) | $\Delta\omega\tau_D$ |
|---|---|---|---|---|---|---|---|---|---|---|
| **Out of motional averaging regime – $\Delta\omega\tau_D$>1 (continued)** | | | | | | | | | | |
| Ai et al [16] | 8 nm SPIO PCL5k-b-PEG5k | Fe₃O₄ cluster | Amphiphilic Copolymer Micelle (PEG-PCL) | 97 | DLS | 0.11 | 41900 | 318 | 1.94·10⁻⁸ | 3.7 |
| | 16 nm SPIO PCL5k-b-PEG5k | | | 110 | | 0.19 | 79200 | 471 | 1.42·10⁻⁸ | 9 |
| Cheong et al [22] | core/shell | Fe–γFe₂O₃ core-shell | Dimercapto-succinic acid (DMSA) | 15.4 | TEM | 1 | 660000 [g] | 324 | 8.3·10⁻¹⁰ | 1.5 |
| Paquet et al [15] | Densely packed SPIONs | γ-Fe₂O₃/Fe₃O₄ mixture cluster | Surfactant (SDS) | 68 | TEM | 0.34 | 98600 | 270 | 9.3·10⁻⁹ | 6 |
| | SPION Cluster core inside a hydrogel shell | γ-Fe₂O₃/Fe₃O₄ mixture cluster | Polymer (PNIPAM) Hydrophilic at T<32°C, Hydrophobic at T>32°C | 88 | | 0.157 | 45500 | 394 | 3·10⁻⁸ | 3.3 |
| | | | | 112 | | 0.076 | 22100 | 420 | 6.6·10⁻⁸ | 2.6 |
| | | | | 130 | | 0.049 | 14100 | 436 | 1.1·10⁻⁷ | 2.2 |
| | | | | 108 | | 0.085 | 24600 | 467 | 6.5·10⁻⁸ | 2.7 |
| | | | | 152 | | 0.03 | 8800 | 484 | 1.9·10⁻⁷ | 1.9 |
| | | | | 176 | | 0.02 | 5700 | 505 | 3.1·10⁻⁷ | 1.6 |
| Jung et al [13] | AMI-25 Feridex Ferumoxide Endorem (brand names) | γ-Fe₂O₃/Fe₃O₄ mixture cluster | Hydrophilic Polymer (Dextran) | 80 | DLS | 0.23 [h] | 77000 | 107 | 4.1·10⁻⁹ | 4.4 |
| Yang et al [23] | Fe₃O₄-MMPNs | Fe₃O₄ cluster | Amphiphilic Copolymer Micelle (PLGA-PEG) | 73 | TEM | 0.118 [i] | 45000 | 333 | 1.92·10⁻⁸ | 2.2 |
| | MnFe₂O₄-MMPNs | MnFe₂O₄ cluster | | 70 | | 0.125 [j] | 51000 | 567 | 2.78·10⁻⁸ | 2.3 |
| Sanson et al [21] | WDi-50 | γ-Fe₂O₃ | Hydrophobic Membrane of Polymersome (PTMC-PGA) | 94 | DLS | 5.8·10⁻² [b] | 16200 | 173 | 3.81·10⁻⁸ | 1.34 |
| | WDi-70 | | | 104 | | 7.1·10⁻² [b] | 19800 | 182 | 3.27·10⁻⁸ | 2.0 |

[g] Core diameter is 9 nm, representing 18% of the particle volume. Alpha-iron density being 7.874 g·cm⁻³, an average mass density of the core-shell of 5.5 g·cm⁻³ was used to derive $M_v$ from the mass magnetization of 115 emu·g⁻¹.
[h] 63.8 wt% iron oxide from thermogravimetry analysis (TGA). The USPIOs have a $d_w$=5.6 nm (TEM, $N$=694).
[i] 40.9 wt% iron ferrite from TGA. The USPIOs have $d_w$=9 nm (TEM, $M_v$=3.3 10⁵ A/m (74 emu·g⁻¹).
[j] 41.7 wt% manganese ferrite from TGA. The USPIOs have $d_w$=9 nm (TEM), $M_v$=4.0·10⁵ A/m (80.8 emu·g⁻¹).

**Table 2.** Saturation magnetizations, optimal diameters and maximum relaxivities for different types of magnetic particles used as $T_2$ MRI contrast agents.

| Materials | $M_v$ [$A \cdot m^{-1}$] [a] | $d_{optimal}$ [nm] [b] | $r_2^{max}$ [$s^{-1} \cdot mM^{-1}$] [c] |
|---|---|---|---|
| Maghemite nanoparticles | $3.5 \times 10^5$ | 55 | 750 |
| Clusters of maghemite nanoparticles at a volume dilution of 5 ($\phi_{intra}$=20% v/v maghemite fraction) | $7 \times 10^4$ | 120 | 750 |
| Iron-iron oxide core-shell nanoparticles | $6.6 \times 10^5$ | 38 | 1200 |
| $(Zn_{0.4}Mn_{0.6})Fe_2O_4$ nanoparticles | $8.75 \times 10^5$ | 35 | 1860 |

[a] saturation magnetization
[b] optimal diameter
[c] maximum transverse relaxivity

**Table of contents entry.** This study evidences size, magnetization and magnetic volume fraction as the only control parameters of MRI $T_2$ contrast agents. Experimental relaxation and magnetometry data on magnetic particles draw up a master curve, allowing the prediction of the efficiency of any nanoparticles or clusters. A calculation of the optimal size for $T_2$ contrast agents of different natures is also performed.

**Keyword list.** Magnetic Resonance Imaging Contrast Agents; Transverse Relaxivity; Motional Averaging Regime; Static Dephasing Regime; Superparamagnetic Iron Oxide Nanoparticles

Quoc L. Vuong, Jean-François Berret, Jérôme Fresnais, Yves Gossuin* and Olivier Sandre*

**Universal Scaling Law to Predict the Efficiency of Magnetic Nanoparticles as MRI T2-Contrast Agents**

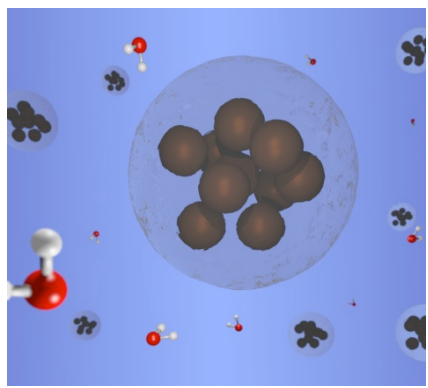

**ToC figure**

# Supporting Information

**Universal Scaling Law to Predict the Efficiency of Magnetic Nanoparticles as MRI T2-Contrast Agents**

By
*Quoc L. Vuong, Jean-François Berret, Jérôme Fresnais, Yves Gossuin\* and Olivier Sandre\**

Dr. Q. L. Vuong, Author-One,
Université de Mons, Biological Physics Department, 20 Place du Parc, 7000 Mons, Belgium
Dr. J.-F. Berret, Author-Two,
Université Denis Diderot Paris-VII, CNRS UMR7057, Matière et Systèmes Complexes
10 rue Alice Domon et Léonie Duquet, 75013 Paris, France
Dr. J. Fresnais, Author-Three,
UPMC Univ Paris 06, CNRS UMR7195, Physicochimie, Colloïdes et Sciences Analytiques
4 place Jussieu, 75005 Paris, France
[*]     Dr. Y. Gossuin, Corresponding-Author,
Université de Mons, Biological Physics Department, 20 Place du Parc, 7000 Mons, Belgium
E-mail: yves.gossuin@umons.ac.be
[*]     Dr. O. Sandre, Corresponding-Author,
Université de Bordeaux, CNRS UMR5629, Laboratoire de Chimie des Polymères Organiques
ENSCBP, 16 Avenue Pey Berland, 33607 Pessac, France
E-mail: olivier.sandre@ipb.fr

Outline:
S1. Synoptic scheme of the size sorting procedure
S2. Characterization of the nanoparticles
    a. Magnetometry
    b. Transmission Electron Microscopy
    c. NMR Relaxometry

## S1. Synoptic scheme of the size sorting procedure

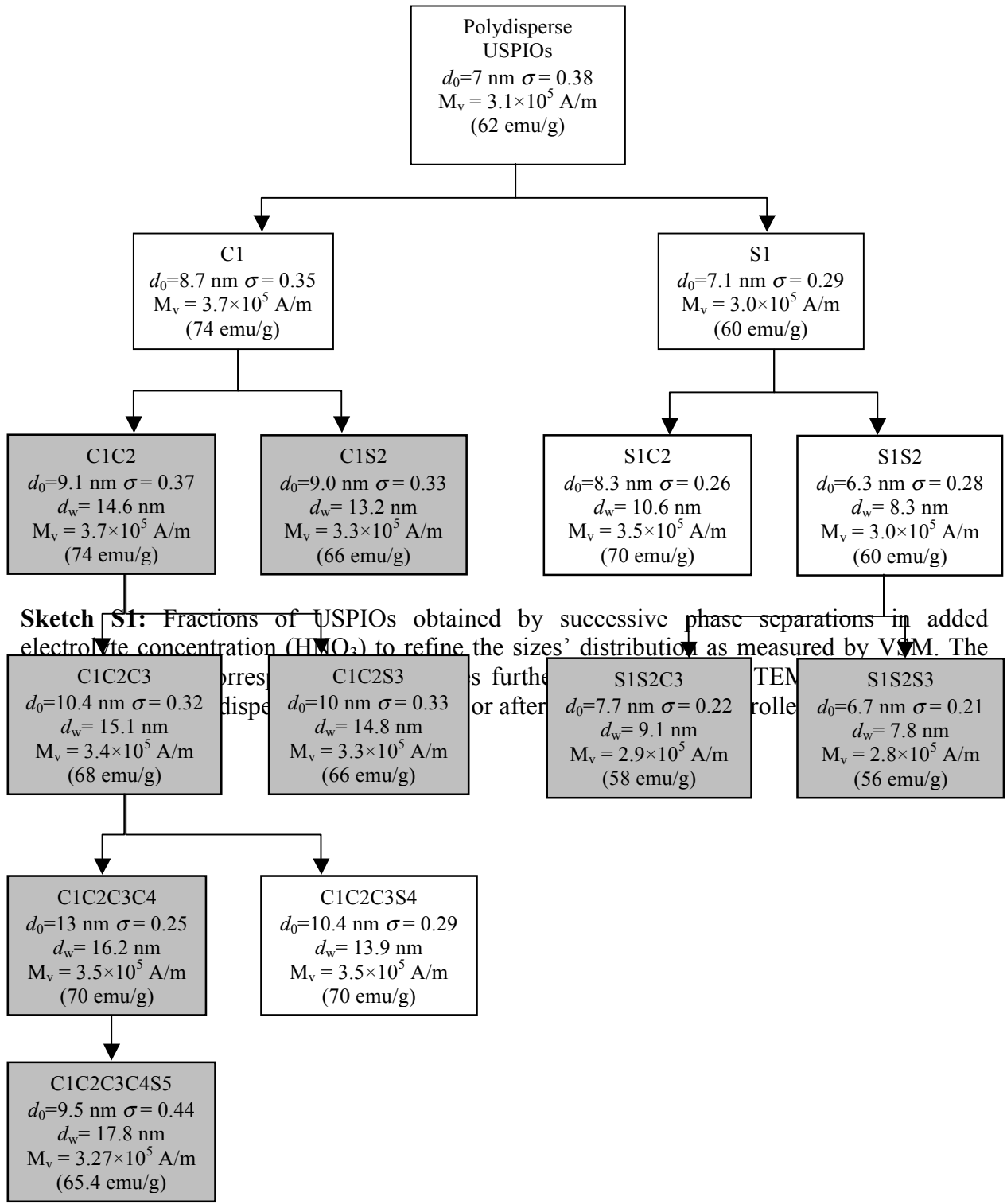

**Sketch S1:** Fractions of USPIOs obtained by successive phase separations in added electrolyte concentration ($HNO_3$) to refine the sizes' distribution as measured by VSM. The [...] corresponding [...] is further [...] dispersed [...] or after [...] TEM [...] controlled

## S2. Characterization of the nanoparticles

### S2.a Magnetometry

We use two populations (S1S2S3 and S1C2) as a comparison. **Figure S1** shows the typical evolution of the macroscopic magnetization $M(H)$ normalized by its saturation value $M_S$ for the γ-Fe$_2$O$_3$ superparamagnetic NP. Here, $M_S = f\,m_S$, where $m_S$ is the specific magnetization of colloidal maghemite (approximately $m_S = 3.5 \times 10^5$ A m$^{-1}$) which is lower than for bulk maghemite. It decreases when the diameter of the superparamagnetic NP decreases due to some disorder of the magnetic moments located near the surface. The solid curves on Figure S1 were obtained by Langevin fits convoluted with log-normal distribution laws of the particle sizes. The parameters of the distribution are the median diameter and the standard variations, respectively $d_0^{VSM} = 6.7 \pm 0.1$nm with $\sigma^{VSM} = 0.21 \pm 0.03$ for S1S2S3 and $d_0^{VSM} = 8.3 \pm 0.1$nm with $\sigma^{VSM} = 0.21 \pm 0.02$ for S1C2.

**Figure S1:** Magnetic field dependence of the macroscopic magnetization $M(H)$ normalized by its saturation value $M_S$ for cationic maghemite dispersions. The solid curve was obtained using the Langevin function for superparamagnetism convoluted with a log-normal distribution function for the particle sizes, given with median diameters $d_0^{VSM}$ and width $\sigma^{VSM}$.

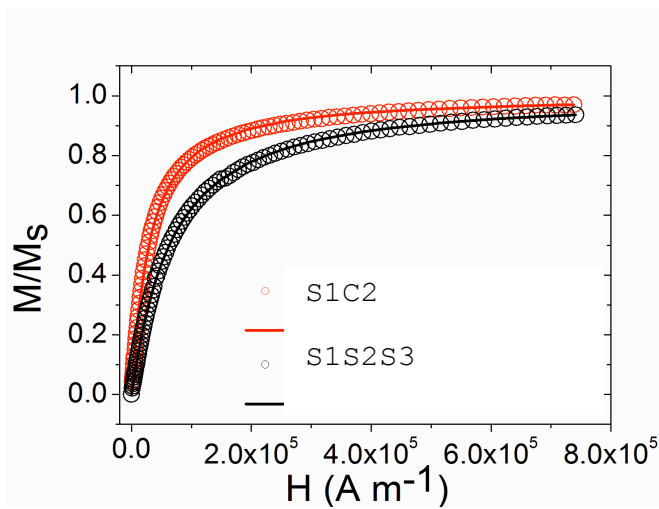

### S2.b Transmission Electron Microscopy (TEM)

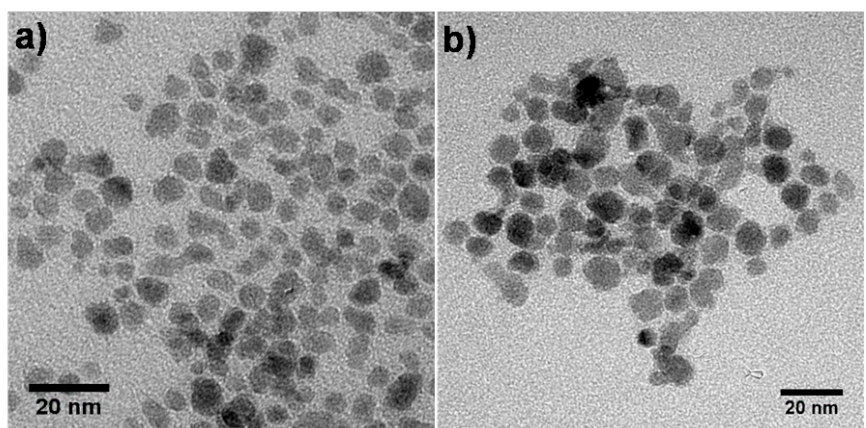

**Figure S2:** Iron oxide superparamagnetic NPs. (a) S1S2S3 and (b) S1C2 as observed by TEM. The stability of the dispersion was ensured by electrostatic interactions mediated by the native cationic charges in diluted HNO$_3$ (pH=1.2– 1.7).

In fact, the median diameter $d_0^{VSM}$ obtained by VSM is related to the crystal structure inside the $\gamma$-Fe$_2$O$_3$ nanoparticle. We then compared these values with the physical diameters $d_0^{TEM}$ by using image analysis of transmission electron microscopy (TEM). **Figure S2** displays images of the two batches of $\gamma$-Fe$_2$O$_3$ USPIOs chosen as examples (S1S2S3 and S1C2).

**Figure S3** shows probability distribution functions of sizes for these NPs observed by TEM on a series of images similar to Figure S2. The data are fitted by a log-normal function with physical diameters $d_0^{TEM}$ = 6.8±0.2 nm and $d_0^{TEM}$ = 9.3±0.2 nm, with polydispersities $\sigma^{TEM}$ = 0.21±0.01 and $\sigma^{TEM}$ = 0.18±0.01.

$$p(d, d_0, \sigma_D) = \frac{1}{\sqrt{2\pi}\sigma_D d} \exp\left(-\frac{\ln^2(d/d_0)}{2(\sigma_D)^2}\right) \qquad (S1)$$

These values are in good agreement with the ones obtained from VSM, albeit with a minor difference between the median diameter $d_0^{VSM}$ and physical diameter $d_0^{TEM}$, which could originate from defects located close to the particles' surface not contributing to the magnetic properties.

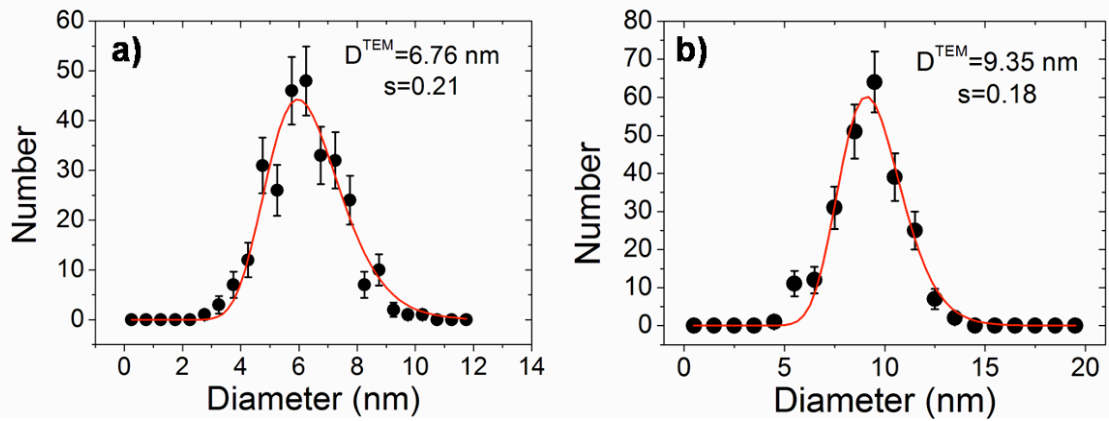

**Figure S3:** Probability distributions function of sizes for two $\gamma$-Fe$_2$O$_3$ USPIOs: a) S1S2S3; b) S1C2. The continuous line was derived from best fit calculation using a Log-normal distribution. For these dispersions, the average diameters by TEM were 6.8 nm and 9.3 nm, and the polydispersity 0.21 and 0.18.

Using the Log-normal distributions deduced by VSM, we can estimate a weight-averaged diameter $d_w = \langle d^4 \rangle / \langle d^3 \rangle$ characteristic of each sample by calculating the 4$^{th}$ and 3$^{th}$ order moments of the Log-normal distributions: $d_w = d_0 \times \exp(3.5 \times \sigma^2)$. This fairly compares to the average diameter $d_{TEM}$ obtained by the analysis of TEM pictures (Figure S3). From this comparison between TEM and VSM, we conclude that the weight average diameter $d_w$ calculated from the two fitting parameters $d_0$ and $\sigma$ of the VSM curve correctly reflects both the size distribution of the samples and their magnetic surface disorder and therefore can be used as a single characteristic size.

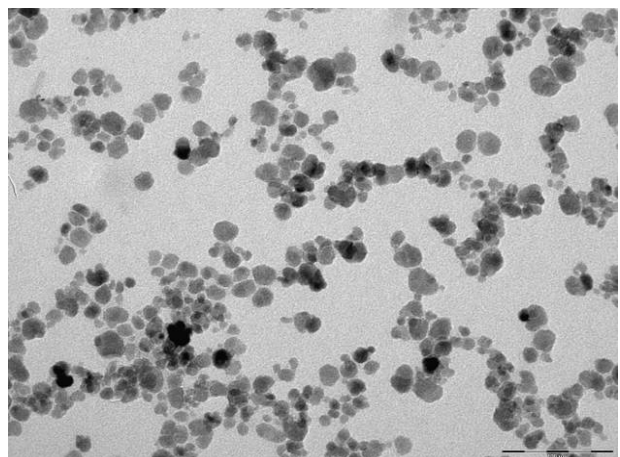

**C1C2C3 in HNO$_3$ (pH 1.5)**

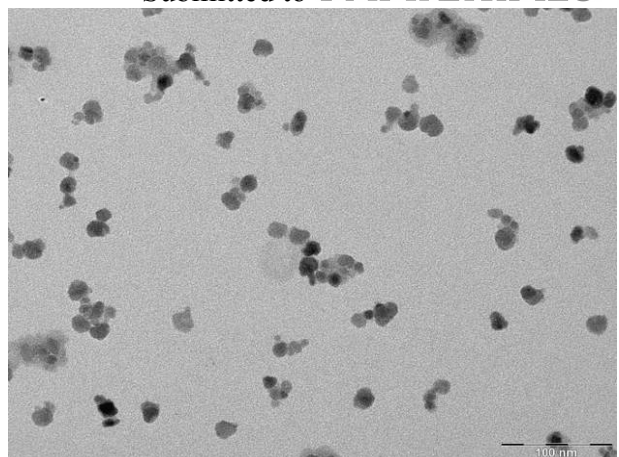

**C1C2C3C4 in HNO$_3$ (pH 1.5)**

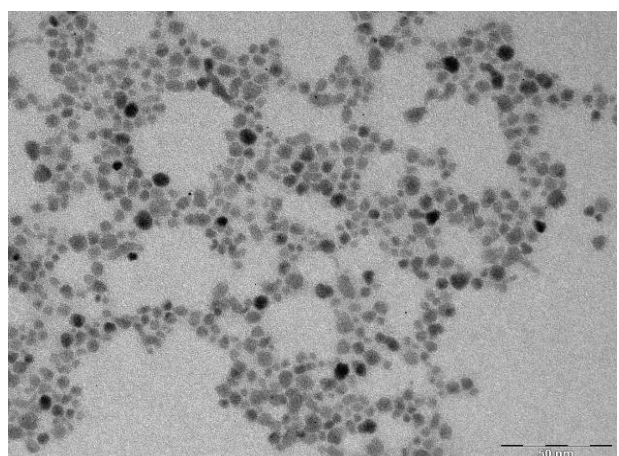

**S1S2-citrate (pH 7)**

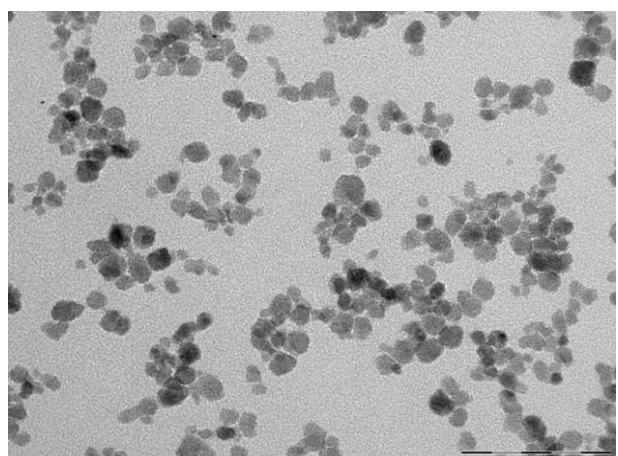

**C1C2S3 in HNO$_3$ (pH 1.5)**

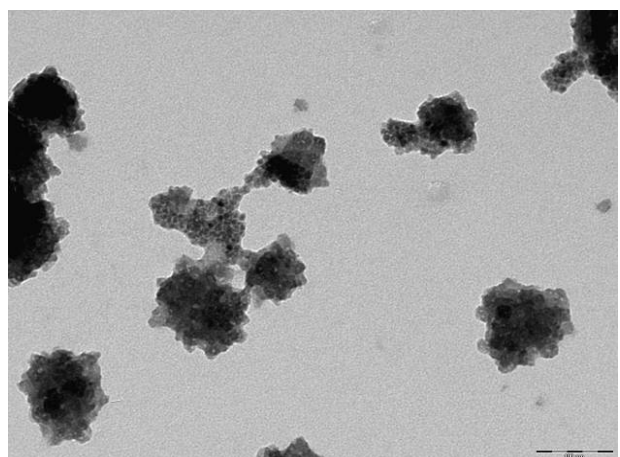

**S1S2C3@PAA/PAM-*b*-PTEA clusters**

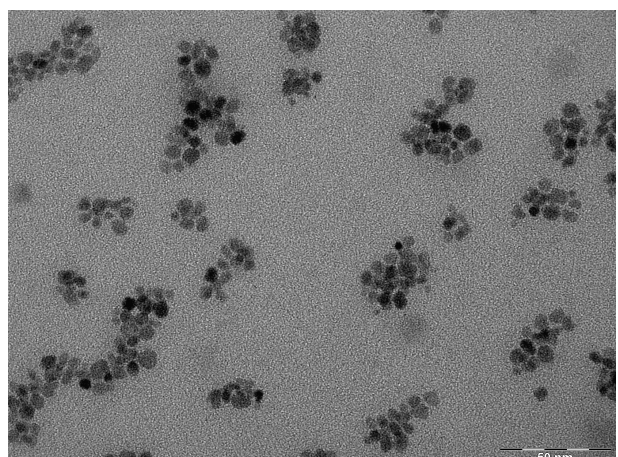

**S1S2S3@PAA/PAM-*b*-PTEA clusters**

**Figure S4**: Typical TEM images of the synthesized particles. The images were acquired on a JEOL100 transmission electron microscope operating at 80 kV by Aude Michel, "Physicochimie des Electrolytes, Colloïdes et Sciences Analytiques" at UPMC Univ Paris 6.

*S2.c NMR Relaxometry*

**Figure S5** shows the longitudinal NMRD profiles for all the (U)SPIO samples prepared in this study. As expected, the $r_1$ relaxivities at high fields are really low.

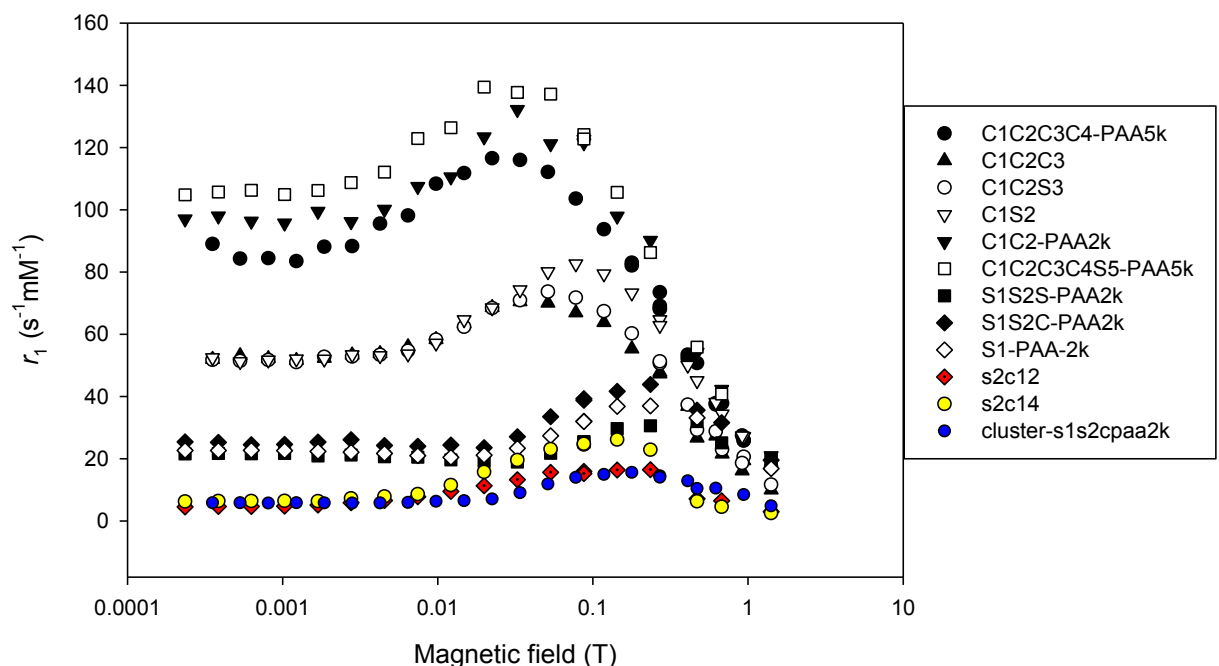

**Figure S5:** longitudinal relaxation of the samples at different magnetic fields.

The transverse relaxivity of all samples at different magnetic fields are shown on **Figure S6**. The relaxivity at 1.41 T was used in the article since it is close to classical clinical imaging fields. We also checked on some samples that the transverse relaxation was almost the same at 1.41 T and at very high fields (*e.g.* 9.4 T/400 MHz), as expected for maghemite particles.

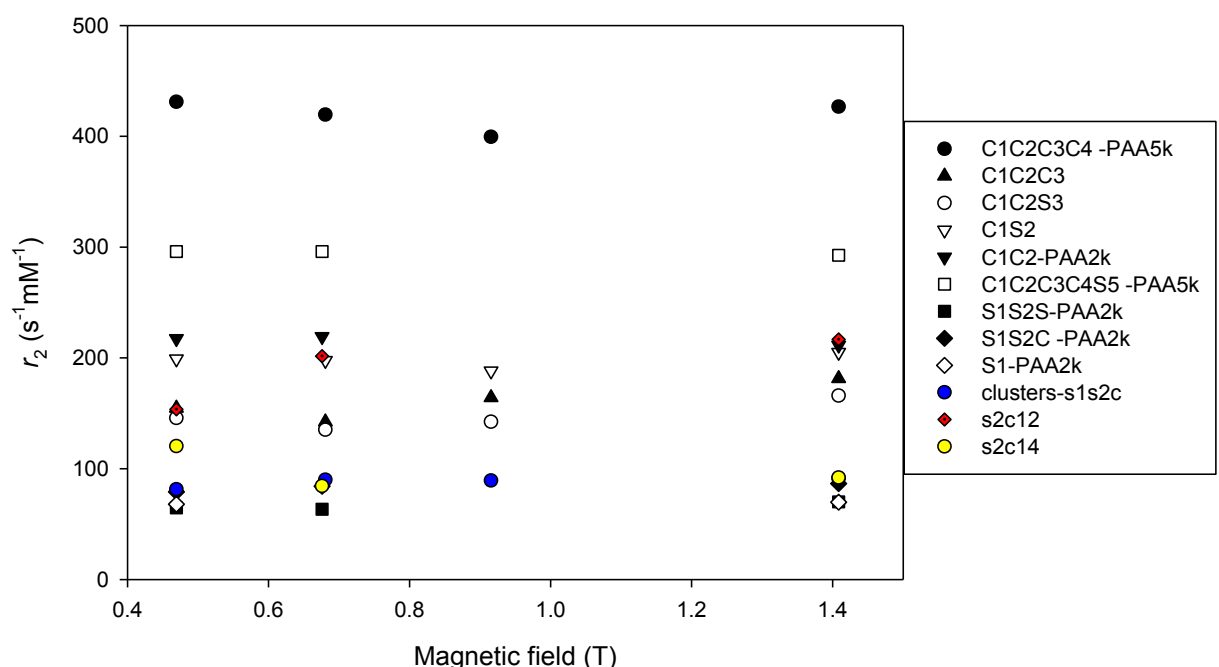

**Figure S6:** transverse relaxation of the samples at different magnetic fields.

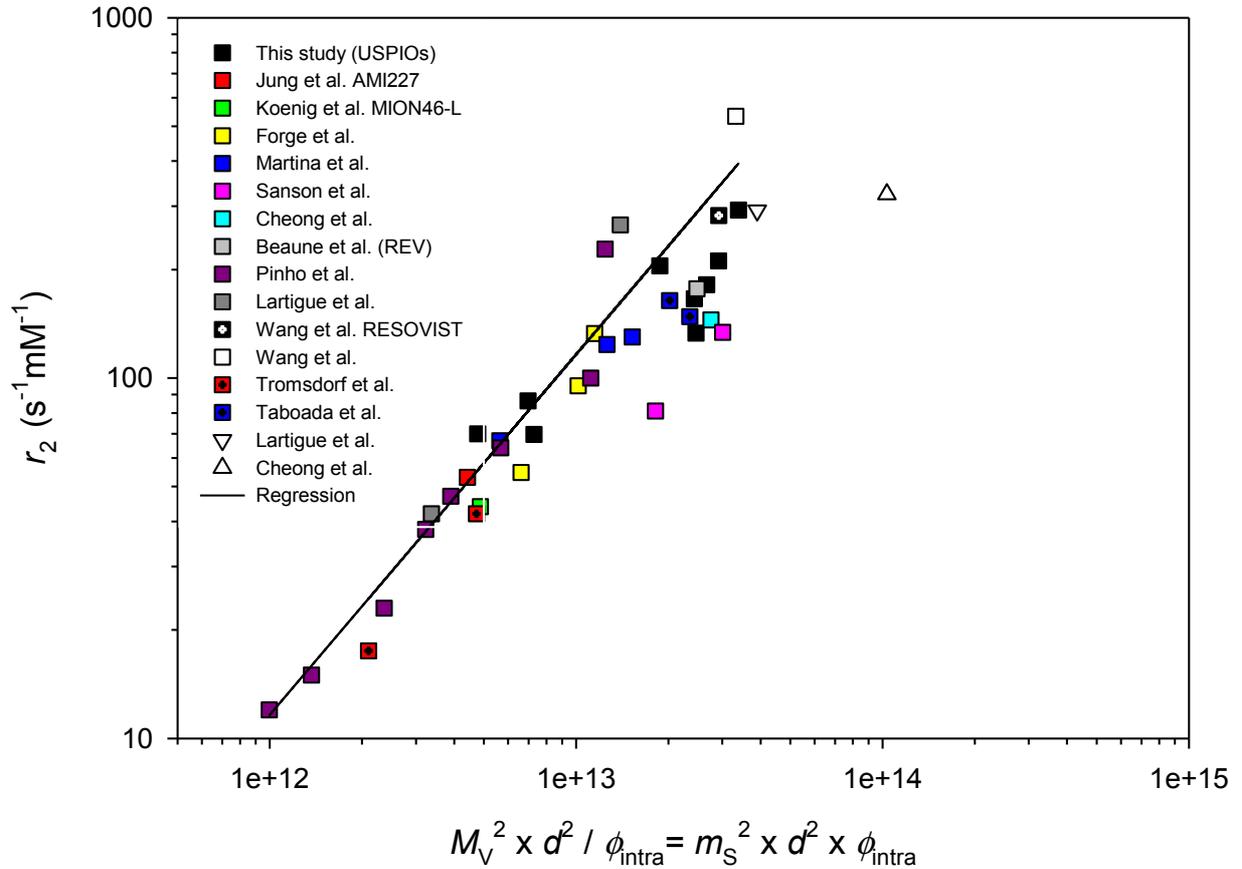

**Figure S7:** Raw values of the transverse relaxivity at high field (≥1T) for samples in the MAR. $r_2$ is appearing linear with the squares of the magnetization and of the diameter divided by the intra-aggregate volume fraction for individual USPIOs (for which $\phi_{intra}$=100%) and clusters (either of low size or dilute) in the MAR *i.e.* satisfying Equation (4).

We can also introduce the specific magnetization of the magnetic cores in the cluster $m_S$ and relate it to the whole body magnetization $M_v=\phi_{intra}\times m_S$. Then the transverse relaxivity $r_2$ becomes linear with $m_S^2\times d^2\times\phi_{intra}$ in the MAR. The other clusters corresponding to $\Delta\omega\tau_D>1$ exhibit a lower $r_2$ than the power law and saturate below a plateau value given by Table 2.